\documentstyle[prl,twocolumn,aps,graphicx]{revtex}
\begin{document}
\bibliographystyle{prsty}
\draft

 \newcommand{\mytitle}[1]{

 \twocolumn[\hsize\textwidth\columnwidth\hsize

 \csname@twocolumnfalse\endcsname #1 \vspace{1mm}]}


\title{Limits on the Curie temperature 
of (III,Mn)V ferromagnetic semiconductors}

\author{John Schliemann$^{1,2}$, J\"urgen K\"onig$^{1,2}$, Hsiu-Hau Lin$^3$,
        and Allan H. MacDonald$^{1,2}$}

\address{$^1$Department of Physics, Indiana University, Bloomington, IN 47405\\
           $^2$Department of Physics, The University of Texas, Austin, TX 78712\\
         $^3$Department of Physics, National Tsing-Hua University, Hsinchu 300,
         Taiwan}
\date{\today}
\maketitle

\begin{abstract}

Mean-field-theory predicts that the Curie temperature $T_c$ of a 
(III,Mn)V ferromagnet will be proportional to the valence
band density-of-states of its host (III,V) semiconductor,
suggesting a route toward room-temperature ferromagnetism in 
this materials class.  In this Letter, we use theoretical 
estimates of spin-wave energies and Monte-Carlo simulations 
to demonstrate that long-wavelength collective fluctuations,
neglected by mean-field theory, 
will limit the critical temperature in large density-of-states
materials.  We discuss implications for high $T_c$ searches.

\end{abstract}

The recent discovery\cite{Ohno99} of ferromagnetism at relatively
high temperatures ($T_c > 100 {\rm K}$) in III-V compound
semiconductors containing Mn has generated intense interest, mainly
because of the new technological roadways that would be opened by
room temperature ferromagnetism in semiconductors with favorable
materials properties.
The search for systems in this materials class with higher critical
temperatures is an important current activity that has been guided thus far
by mean-field theoretical\cite{dietl_mft,us_mft} considerations.
In this Letter we address the importance of collective
magnetization fluctuations, neglected by mean-field theory, in
limiting the critical-temperature and discuss the implications of
these considerations for high-$T_c$ searches. 

Our analysis is based on the kinetic-exchange model for interactions between 
the Mn spins and band electrons,
\begin{equation}
\label{model}
   H = H_0 + J_{pd} \sum_I \int d^3 r \,
	\vec S_I \cdot \vec s (\vec r) \,\delta(\vec r - \vec R_I)\, ,
\end{equation}
where $\vec S_I$ describes a Mn spin with spin length $S=5/2$
at site $\vec R_I$, $\vec s(\vec r)$ is
the band-carrier density, and $J_{pd}>0$ represents the exchange integral.
$H_{0}$ represents a simplified single-parabolic-band model for the host 
semiconductor valence bands.

The simplest treatment of this model is a mean field theory which takes
the magnetizations of carriers and ion spins to be uniform in space and
neglects correlations between them. A straightforward calculation yields
for the critical temperature in mean field approximation\cite{dietl_mft,us_mft}
\begin{equation}
T_c^{MF} = \frac{\chi_{P}}
{(g^{\ast} \mu_B/2)^2} \frac{S(S+1)NJ_{pd}^{2}}{12}\,,
\label{tc_explicit}
\end{equation}
where $g^{\ast}$ is the g-factor of the carriers and 
$\chi_{P}$ is their Pauli susceptibility, which is proportional to the
effective band mass. This observation has given rise to concrete predictions
for critical temperatures for several host semiconductors based on their
different band masses \cite{dietl_mft}.

In mean-field theory, ferromagnetism occurs because the
penalty in entropic free energy paid to polarize the Mn spins
vanishes at $T=0$. Any coupling to a band-electron system with a finite
magnetic susceptibility is sufficient to yield ferromagnetism.
While this mean-field theory probably captures much of the physics
of (III,Mn)V ferromagnetism, it has a qualitative deficiency which
will have an important quantitative impact on $T_c$ predictions in
circumstances we identify below. Mean-field theory fails to account
for the small energy cost of magnetization configurations in which
spin orientations vary slowly\cite{coursegrain} in space, reducing the average
magnetization but maintaining local correlations between Mn and
band-electrons spin orientations. 
In the following paragraphs we estimate the critical temperature for the
case when these collective excitations dominate thermal magnetization
suppression.

Isotropic ferromagnets\cite{anisotropy} have spin-wave Goldstone
collective modes whose energies vanish at long wavelengths,
\begin{equation}
\label{long-wavelength}
   E(k) = D k^2 + \cdots \, ,
\end{equation}
where $k$ is the wavevector of the mode. Each spin-wave excitation
reduces the total spin of the ferromagnetic state by 1. The
coefficient $D$ is inversely proportional to the saturation 
magnetization and proportional to the exchange constant 
$A$ of classical micromagnetic theory,
that parameterizes the free-energy cost of spatial variations in
magnetization orientation.  We have previously presented a
theory of spin-wave excitations in (III,Mn)V ferromagnets
\cite{us_prl}. These collective excitations are not accounted for
in the mean-field approximation. If the spin stiffness is small,
they will dominate the suppression of the magnetization at all
finite temperatures and limit the critical temperature. 
A rough upper bound \cite{rough} on the resulting critical temperature, 
$T_c^{coll}$, can be found by using the $T=0$ stiffness value and finding the
temperature where the number of excited spin waves per unit volume
equals the spin per volume of the ground state,
\begin{equation}
   SN = {1\over 2 \pi^2} \int_0^{k_D} dk\, k^2 n[E(k)] \, .
\end{equation}
A Debye wavevector\cite{coursegrain} $k_D = (6\pi^2 N)^{1/3}$ cuts
off the sum over wavevectors at the correct number of modes, and 
$n(E)$ is the Bose occupation number.
We find that
\begin{equation}
   k_BT_c^{coll} = {2S+1\over 6} D k_D^2
\label{collective_tc}
\end{equation}
for $S\ge 5/2$.
To obtain this equation, we have assumed that the spin waves can be
approximated as non-interacting Bose particles, replaced the dispersion by the 
long-wavelength limit Eq.~(\ref{long-wavelength}), and noted that the
critical temperature estimate is proportional to $D k_D^2$,
justifying the use of the classical expression for the mode
occupation number $n(k) \approx k_B T/E(k)-1/2$. These considerations
set an upper bound on the critical temperature which is proportional to
the spin-stiffness, a bound not respected by the mean-field theory.
A familar example of ferromagents in which long range order is suppressed 
by long-wavelength collective excitations is provided by the 
ferromagnetic transition metals Fe, Co, and Ni.  In that case,
an expression similar to Eq.~(\ref{collective_tc}) and proportional
to the micromagnetic exchange constant, predicts critical temperatures
with $20 \%$ accuracy\cite{lp,Gunnar} whereas mean-field-theory 
overestimates $T_c$ 
by factors of 5 to 10.  In (III,Mn)V ferromagnets, we will see that 
both mean-field and 
collective regimes can occur depending on carrier density and host
semiconductor band parameters.
\begin{figure}
\centerline{\includegraphics[width=8cm]{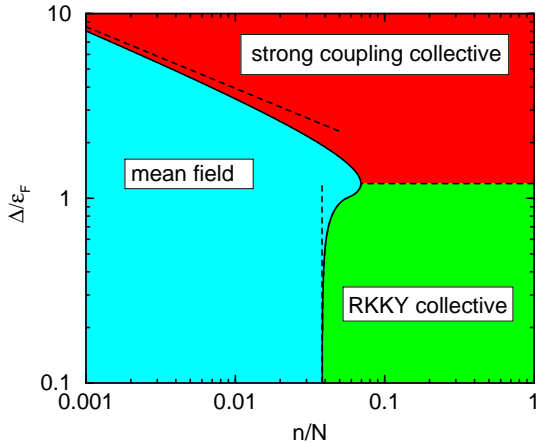}}
\caption{Critical-temperature-limit regimes.  In the mean-field regime $T_c$ is
limited by individual Mn spin fluctuations. In the collective
regimes, the critical temperature is limited by long-wavelength
fluctuations with a stiffness proportional to the bandwidth for
weak (RKKY) exchange coupling and inversely proportional to the
bandwidth for strong exchange coupling. At the solid line
$T_c^{MF}=T_c^{coll}$. Dashed lines: expansions for large and small
$\Delta/\epsilon_F$, Eqs.~(\ref{coll_RKKY}) and (\ref{coll_s}), and the 
crossover from the RKKY to the strong coupling collective regime. }
\label{fig1}
\end{figure}
Our theoretical results for (III,Mn)V ferromagnet
collective modes lead to physically transparent results for the
spin-stiffness in both strong and weak exchange coupling limits.
The dimensionless parameter which characterizes the strength of the
exchange coupling is the ratio $\Delta/\epsilon_F$, where
$\epsilon_F$ is the band-system Fermi energy and $\Delta = J_{pd} N
S$ is its mean-field spin splitting at $T=0$. For small
$\Delta/\epsilon_F$, the RKKY regime\cite{dietl_mft},
exchange coupling is a weak
perturbation on the band system. In this regime we
find that $D = \delta/(12k_F^2)$ where $\delta =
J_{pd} n \xi/2 = (3/8S)(n/N)(\Delta^2/\epsilon_F)$ is the energy
cost of an uncorrelated spin reduction at a single Mn site. Note
that in this regime $\delta \sim T_c^{MF}$ and that 
\begin{equation}
\label{coll_RKKY}
   T_c^{coll,RKKY} = T_c^{MF} {2S+1 \over 12(S+1)\sqrt[3]{2}}
	\left({N\over n}\right)^{2/3} \, .
\end{equation}
In the weak coupling regime mean-field theory is reliable only for $n/N \ll 1$,
as expected since in this case the RKKY interaction has a range
which is long compared to the distance between Mn spins. In the
large $\Delta/\epsilon_F$ regime exchange coupling completely
polarizes the band-electron system. In this case (and for $n \ll
2NS$) we find that $D = (n/2NS)( \epsilon_F/k_F^2)$.
For a fully polarized band the energetic cost of varying the moment
orientation direction is entirely due to band kinetic energy. We,
thus, obtain as a third $T_c$ bound
\begin{equation}
\label{coll_s}
   T_c^{coll,s} = {2S+1\over 12S}\epsilon_F \left({n\over N}\right)^{1/3}\, . 
\end{equation}
The different regimes deduced from these considerations are illustrated in 
Fig.~\ref{fig1}.

To substantiate these qualitative considerations, we have performed
Hybrid-Monte-Carlo \cite{HMC} simulations, treating the Mn spins as
discrete classical degrees of freedom, an approximation that is justified
near the critical temperature. We allow for disorder by choosing
the Mn positions randomly. Microscopic $p-d$ exchange physics is
modeled by allowing the interaction to have a finite range $a_0$
\cite{Batta}. We simulate this by replacing the delta function by a Gaussian
distribution in Eq.~(\ref{model}).

An exhaustive description  of our Monte Carlo approach, including a detailed 
account of all technical aspects such as thermalization procedures, finite-size
effctes etc., will be given elsewhere \cite{lp}. 
Here we shall, for brevity, concentrate on the results. 
In the following we consider the strong
coupling regime where mean-field theory is not reliable and
finite-size effects in our simulations are small. One important
finding is that randomness in the Mn positions can {\em enhance} the spin
stiffness (for large $J_{pd}$ by up to factor of two) in
comparison to the estimates\cite{us_prl} based on course-grained Mn
spins with Debye cutoffs discussed above. The importance of disorder was
also emphasized by Wan and Bhatt \cite{WaBh:00}
for a model of interacting magnetic polarons
where both the ion and the carrier spins are treated classically,
while we account for the quantum-mechanical nature of the free
carriers in our simulation.

Figure \ref{fig2} shows a typical magnetization curve for manganese
ions and carriers magnetizations in the collective regime. The
maximum of the finite-size Mn susceptibility marks the
ferromagnetic transition temperature. The lower inset compares the
{\em global} and {\em local} free carrier polarizations,
$m_{loc}=\overline{\left\langle {|\vec s(\vec r)|}/{n(\vec r)}
\right\rangle}$ .
As anticipated by our qualitative discussion, the typical free
carrier local band polarization remains large {\em above} the critical
temperature; it would vanish in a mean-field picture.
Ferromagnetism, and the technologically useful robust collective
physics it gives rise to, disappears in this case only because of
the loss of long-range spatial coherence.
\begin{figure}
\centerline{\includegraphics[width=8cm]{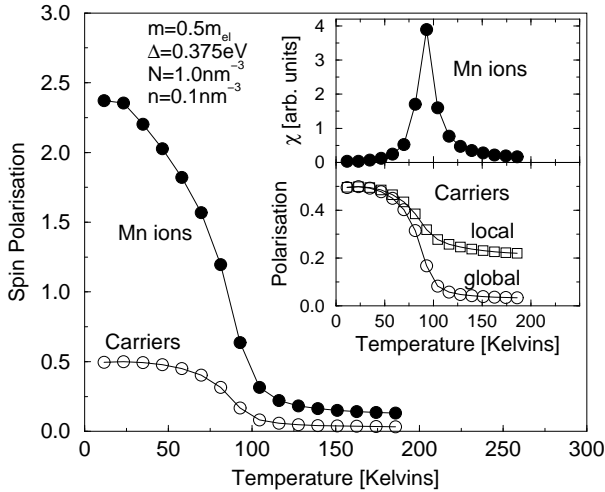}}
\caption{Magnetization curves for Mn ions and carriers. The upper inset
shows the magnetic susceptibility due to the ions, while in the
lower inset the local and the global spin polarization per carrier
is plotted. The data was obtained for a system of 540 Mn spins and 
54 carriers.}
\label{fig2}
\end{figure}
\begin{figure}
\centerline{\includegraphics[width=8cm]{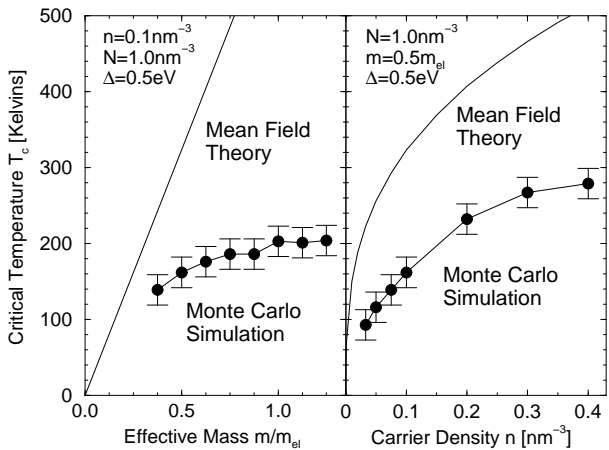}}
\caption{The critical temperature $T_{c}$ as a function of the carrier mass
(left panel) and as a function of the carrier density (right panel).
The results of the Monte Carlo runs are compared with the mean-field
predictions.
}
\label{fig3}
\end{figure}
The left panel of
Fig.~\ref{fig3} shows the critical temperature as a function of
the carrier effective mass.  Mean-field theory predicts that $T_{c}$
will grow linearly with increasing mass. The Monte-Carlo results,
however, are substantially lower and show a saturation of $T_{c}$
at carrier masses close to the bare electron mass. For even higher
masses, we expect $T_{c}$ to decrease, reflecting the reduction of
spin stiffness expected in this regime.

The right panel of
Fig.~\ref{fig3} compares the mean-field prediction and the Monte
Carlo results for $T_{c}$ as a function of the carrier density. For
higher carrier densities we expect ferromagnetism to give way to
spin-glass order.

We conclude from the present work that high critical temperatures cannot
be achieved simply by narrowing the free carrier band or 
placing its Fermi energy at a density-of-states peak in order 
to enhance its Pauli magnetic susceptibility $\chi_P$. 
It will also be necessary
to engineer a suppression of collective magnetization fluctuations.

We thank Glenn Martyna for pointing out to us the striking advantages of
the Hybrid Monte Carlo algorithm, and T. Dietl, T. Jungwirth and 
H. Ohno for useful discussions. We acknowledge support from the Deutsche
Forschungsgemeinschaft, the National Science Foundation, and the Indiana
21st Century Fund.


\begin{thebibliography}{10}
\bibitem{coursegrain} For these qualitative considerations we 
replace the discrete Mn spins by a continuous distribution.  This
is a good approximation when the Mn 
density is much larger than the free-carrier density.
\bibitem{rough}
At finite temperature interactions between spin waves will reduce the spin 
stiffness. To model this effect a self-consistent spin-wave scheme would be
required.
Here, however, we use the zero-temperature spin stiffness for an upper 
bound.
\end{thebibliography}
\end{document}